\documentclass{aa}
\usepackage{graphicx}
 
\begin{document}

\title{Detection of 75+ pulsation frequencies in the $\delta$ Scuti star FG Vir}
 
\author{M.~Breger\inst{1} \and P.~Lenz\inst{1} \and V.~Antoci\inst{1}
\and E.~Guggenberger\inst{1}
\and R.~R.~Shobbrook\inst{2}
\and G.~Handler\inst{1}
\and B.~Ngwato\inst{3}
\and F.~Rodler\inst{1}
\and E.~Rodriguez\inst{4}
\and P. L\'opez de Coca\inst{4}
\and A. Rolland \inst{4}
\and V.~Costa \inst{4}}

\institute{Institut f\"ur Astronomie der Universit\"at Wien, T\"urkenschanzstr. 17,
A--1180 Wien, Austria\\
e-mail: michel.breger@univie.ac.at
\and
Research School of Astronomy and Astrophysics, Australian National University, Canberra, ACT, Australia
\and
Theoretical Astrophysics Programme, University of the North-West,
Private Bag X2046, Mmabatho 2735, South Africa
\and
Instituto de Astrofisica de Andalucia, CSIC, Apdo. 3004, E-18080 Granada, Spain}

\date{Received date; accepted date}

\abstract{ Extensive photometric multisite campaigns of the $\delta$~Scuti variable FG~Vir are presented.
For the years 2003 and 2004, 926 hours of photometry at the millimag precision level were obtained. The combinations
with earlier campaigns lead to excellent frequency resolution and high signal/noise. A multifrequency
analysis yields 79 frequencies. This represents a new record for this type of star. The modes
discovered earlier were confirmed.

Pulsation occurs over a wide frequency band from 5.7 to 44.3 c/d with amplitudes of 0.2 mmag or larger.
Within this wide band the frequencies
are not distributed at random, but tend to cluster in groups. A similar
feature is seen in the power spectrum of the residuals after 79 frequencies are prewhitened. This indicates that
many additional modes are excited. The interpretation is supported by a histogram
of the photometric amplitudes, which shows an increase of modes with small amplitudes. The old question of
the 'missing modes' may be answered now: the large number of detected frequencies as well as
the large number of additional frequencies suggested by the power spectrum
of the residuals confirms the theoretical prediction of a large number of excited modes.

FG~Vir shows a number of frequency combinations of the dominant mode at 12.7162 c/d ($m$ = 0) with
other modes of relatively high photometric amplitudes. The amplitudes of the frequency sums are higher than those of the differences.
A second mode (20.2878 c/d) also shows combinations. This mode of azimuthal order $m$ = -1 is coupled with two other modes of $m$ = +1.
}
\maketitle
\section{Introduction}

The $\delta$ Scuti variables are stars of spectral type A and F in the main-sequence or
immediate post-main-sequence stage of evolution. They generally pulsate with a large number of simultaneously excited radial
and nonradial modes, which makes them well-suited for asteroseismological studies. The
photometric amplitudes of the dominant modes in the typical $\delta$ Scuti star are a few millimag.
It is now possible for ground-based telescopes to detect a large number of simultaneously excited modes
with submillimag amplitudes in stars other than the Sun (e.g., Breger et al. 2002, Frandsen et al. 2001).
Because photometric studies measure the integrated light
across the stellar surface, they can detect low-degree modes only. This is a simplification
for the interpretation because of fewer possibilities in mode identification.

A typical multisite photometric campaign allows the discovery of about five to ten frequencies
of pulsation from about 200 to 300 hours of high-precision photometry (e.~g., V351 Ori, Ripepi et al. 2003;
V534 Tau, Li et al. 2004). These excellent observational studies are then compared to theoretical pulsation
models, but the fit is hardly unique (e.g., $\theta^2$ Tau, Breger et al. 2002b). The uniqueness problem can be lessened
by studies with even lower noise in the power spectrum. This can be achieved by the very accurate measurements from space and
by larger ground-based studies with more data, which concentrate on a single selected star. These more extensive
ground-based studies also lead to to higher frequency resolution. The latter is important, because
$\delta$~Scuti stars can show a large number of very close frequency pairs (or groups), which can only be resolved
through long-term studies lasting many months or years. The question of frequency resolution is an important aspect
in planning asteroseismological space missions (e.g., see Handler 2004, Garrido \& Poretti 2004).

The Delta Scuti Network (DSN) is a network of telescopes situated on different continents. The collaboration reduces
the effects of regular daytime observing gaps. The network
 is engaged in a long-term program (1000+ hours of observation, 10+ years,
photometry and spectroscopy) to determine the structure and nature of the multiple frequencies
of selected $\delta$ Scuti stars situated in different
parts of the classical instability strip. The star FG~Vir is the present main long-term target
of the network. This 7500K star (Mantegazza, Poretti \& Bossi 1994) is at the end of its main-sequence evolution.
The projected rotational velocity is very small (21.3 $\pm$ 1.0 km s$^{-1}$, Mittermayer \& Weiss 2003, see also
Mantegazza \& Poretti 2002).

A number of photometric studies of the variability of FG Vir are available:
a lower-accuracy study by Dawson from the years 1985 and 1986
(Dawson, Breger \& L\'opez de Coca 1995, data not used here), high-accuracy studies from
1992 (Mantegazza, Poretti \& Bossi 1994), as
well as previous campaigns by the Delta Scuti Network in 1993, 1995 and 2002 (Breger et al. 1995, 1998, 2004).
Furthermore, for the year 1996, additional $uvby$ photometry is available (Viskum et al. 1998). 12 nights
of data were of high accuracy and could be included.

Because of the large scope of the long-term project on the pulsation of FG~Vir, the photometric, spectroscopic
and pulsation-model results cannot be presented in one paper. Here we present the extensive new photometric data
from 2003 and 2004 as well as multifrequency analyses to extract the multiple frequencies excited in FG~Vir. The
analyses concentrate on the available three years of extensive coverage (2002--2004) and also consider
the previous data (1992--1996).

Separate studies, presently in progress, will (i) present mode identifications based mainly on high-dispersion
line-profile analyses and the data presented in this paper, (ii) examine the nature of close frequencies, 
(iii) compute asteroseismological models of stellar structure to fit the observed frequency spectrum.

\section{New photometric measurements}

During 2003 and 2004, photometric measurements of the star FG~Vir were scheduled for
$\sim350$ nights at four observatories. Of these, 218 nights were of high photometric
quality at the millimag level with no instrumental problems. These are listed in Table 1 together with the
additional details:

\begin{enumerate}
\item The APT measurements were obtained with the T6 0.75~m Vienna Automatic Photoelectric Telescope
(APT), situated at Washington Camp in Arizona (Strassmeier et al. 1997, Breger \& Hiesberger 1999).
The telescope has been used before for several lengthy campaigns of the Delta Scuti Network, which confirmed
the long-term stability and millimag precision of the APT photometry.

\item The OSN measurements were obtained with the 0.90~m telescope located
at 2900m above sea level in the South-East of Spain at the Observatorio
de Sierra Nevada in Granada, Spain. The telescope is equipped with a simultaneous four-channel
photometer ($uvby$ Str\"omgren photoelectric photometer). The observers for 2003 were:
E.~Rodriguez, P. L\'opez de Coca, A. Rolland, and V.~Costa.

\item The SAAO measurements were made with the Modular Photometer
attached to the 0.5~m and the UCT photometer attached to the
0.75~m telescopes of the South African Astronomical Observatory. The observers were
V.~Antoci, E.~Guggenberger, G.~Handler and B.~Ngwato.

\item The 0.6-m reflector at Siding Spring Observatory, Australia, was used with a
PMT detector. The observers were P. Lenz and R. R. Shobbrook.

\end{enumerate}

The measurements were made with Str\"omgren $v$ and $y$ filters. Since telescopes and photometers
at different observatories have different zero-points, the measurements need to be adjusted. This was done by
zeroing the average magnitude of FG~Vir from each site and later readjusting the zero-points by using the
final multifrequency solution. The  shifts were in the submillimag range. We also checked
for potential differences in the effective wavelength at different observatories
by computing and comparing the amplitudes of the dominant mode. No problems were found.

The measurements of FG~Vir were alternated with those of two comparison stars. Details on the
three-star technique can be found in Breger (1993). We used the same comparison stars as during the
previous DSN campaigns of FG Vir, viz., C1 = HD 106952~(F8V) and C2 = HD~105912 (F5V).
No variability of these comparison stars was found. The two comparison stars also make
it possible to check the precision of the different observing sites. The residuals from the assumed constancy
were quite similar,
i.e., for the (C1--C2) difference we find a standard deviation of $\pm$ 3 mmag for all
observatories and passbands except for $\pm$ 2 mmag  (2004 SAAO75 $v$ as well as $y$ passbands)
and $ \pm$ 4 mmag (2004 APT75 $v$ and 2003 OSN90 $y$ measurements. The power spectrum of the C1--C2 differences
does not reveal any statistically significant peaks.

\begin{table*}
\caption[]{Journal of the PMT observations of FG Vir for 2003 and 2004}
\begin{flushleft}
\begin{tabular}{lll|lll|lll|lll}
\hline
\noalign{\smallskip}
Start & Length & Obs./& Start & Length & Obs./&Start & Length & Obs./&Start & Length & Obs./\\
HJD & hours & Tel. & HJD & hours& Tel. & HJD & hours & Tel. & HJD & hours & Tel. \\
%245 0000+ & &Telescope& 245 0000+& & Telescope & 245 0000+ & &Telescope\\
\noalign{\smallskip}
\hline
\noalign{\smallskip}
\multicolumn{3}{l}{245 000+\hspace{12mm}Year 2003} & 2748.29 & 4.6 & SAAO50 & 2788.89 & 4.5 & SSO60 & 3108.63 & 4.8 & APT75 \\
2656.99 & 1.7 & APT75 & 2748.64 & 6.5 & APT75 & 2792.66 & 2.9 & APT75 & 3109.62 & 6.9 & APT75 \\
2667.81 & 5.7 & APT75 & 2748.89 & 4.9 & SSO60 & 2796.66 & 2.7 & APT75 & 3110.26 & 2.0 & SAAO50 \\
2668.91 & 3.6 & APT75 & 2749.25 & 5.9 & SAAO50 & 2798.67 & 2.3 & APT75 & 3110.63 & 6.9 & APT75 \\
2670.81 & 5.8 & APT75 & 2749.64 & 6.5 & APT75 & 2802.66 & 2.4 & APT75 & 3111.63 & 6.8 & APT75 \\
2671.64 & 0.7 & OSN90 & 2749.89 & 6.7 & SSO60 & 2803.66 & 2.3 & APT75 & 3113.23 & 6.5 & SAAO50 \\
2673.79 & 6.1 & APT75 & 2750.26 & 5.5 & SAAO50 & 2805.66 & 2.2 & APT75 & 3114.64 & 5.3 & APT75 \\
2674.61 & 3.4 & OSN90 & 2750.70 & 5.0 & APT75 & 2806.66 & 2.0 & APT75 & 3115.63 & 6.5 & APT75 \\
2675.62 & 3.1 & OSN90 & 2750.93 & 5.8 & SSO60 & 2809.88 & 2.8 & SSO60 & 3117.63 & 5.9 & APT75 \\
2676.63 & 3.0 & OSN90 & 2751.36 & 2.9 & OSN90 & 2812.66 & 1.6 & APT75 & 3118.63 & 6.3 & APT75 \\
2677.78 & 4.4 & APT75 & 2751.37 & 2.3 & SAAO50 & 2813.66 & 1.8 & APT75 & 3119.63 & 6.1 & APT75 \\
2686.81 & 0.7 & APT75 & 2751.91 & 2.8 & SSO60 & 2813.87 & 1.8 & SSO60 & 3120.63 & 6.0 & APT75 \\
2692.75 & 6.0 & APT75 & 2752.90 & 3.4 & SSO60 & 2814.65 & 0.8 & APT75 & 3123.74 & 3.2 & APT75 \\
2693.82 & 4.2 & APT75 & 2753.24 & 6.0 & SAAO50 & 2816.66 & 1.4 & APT75 & 3125.63 & 5.8 & APT75 \\
2699.96 & 1.9 & APT75 & 2753.64 & 6.2 & APT75 & \multicolumn{3}{c}{Year 2004}& 3130.64 & 5.1 & APT75 \\
2703.77 & 1.8 & APT75 & 2754.25 & 5.7 & SAAO50 & 3022.83 & 5.3 & APT75 & 3131.65 & 4.8 & APT75 \\
2706.71 & 7.3 & APT75 & 2754.64 & 6.0 & APT75 & 3023.83 & 5.0 & APT75 & 3132.66 & 4.6 & APT75 \\
2707.72 & 1.5 & APT75 & 2754.90 & 4.4 & SSO60 & 3031.81 & 1.3 & APT75 & 3136.64 & 4.8 & APT75 \\
2709.71 & 7.4 & APT75 & 2755.25 & 5.4 & SAAO50 & 3032.81 & 5.8 & APT75 & 3137.25 & 3.1 & SAAO50 \\
2710.71 & 7.3 & APT75 & 2755.66 & 5.5 & APT75 & 3033.93 & 2.8 & APT75 & 3137.64 & 4.7 & APT75 \\
2711.71 & 6.1 & APT75 & 2756.25 & 5.5 & SAAO50 & 3034.85 & 4.9 & APT75 & 3138.23 & 2.7 & SAAO50 \\
2712.67 & 1.2 & OSN90 & 2757.65 & 5.6 & APT75 & 3035.82 & 5.1 & APT75 & 3138.64 & 4.4 & APT75 \\
2713.70 & 4.0 & APT75 & 2758.64 & 3.9 & APT75 & 3049.98 & 1.6 & APT75 & 3139.24 & 3.1 & SAAO50 \\
2714.44 & 3.1 & OSN90 & 2758.95 & 4.6 & SSO60 & 3051.00 & 1.0 & APT75 & 3139.64 & 4.4 & APT75 \\
2719.77 & 5.3 & APT75 & 2759.65 & 5.6 & APT75 & 3051.80 & 5.9 & APT75 & 3140.22 & 3.2 & SAAO50 \\
2720.02 & 2.4 & SSO60 & 2759.88 & 4.8 & SSO60 & 3052.76 & 5.9 & APT75 & 3141.64 & 1.9 & APT75 \\
2720.81 & 4.2 & APT75 & 2760.37 & 3.6 & OSN90 & 3053.90 & 0.8 & APT75 & 3142.22 & 3.1 & SAAO50 \\
2720.98 & 6.1 & SSO60 & 2760.64 & 5.6 & APT75 & 3057.74 & 7.2 & APT75 & 3144.23 & 3.9 & SAAO75 \\
2721.68 & 7.4 & APT75 & 2760.93 & 5.4 & SSO60 & 3060.76 & 6.6 & APT75 & 3145.74 & 1.6 & APT75 \\
2722.67 & 7.4 & APT75 & 2761.72 & 3.7 & APT75 & 3061.73 & 7.4 & APT75 & 3146.20 & 1.7 & SAAO75 \\
2722.98 & 5.7 & SSO60 & 2761.88 & 4.6 & SSO60 & 3062.90 & 0.9 & APT75 & 3146.65 & 3.3 & APT75 \\
2723.75 & 4.3 & APT75 & 2762.36 & 4.1 & OSN90 & 3064.73 & 6.8 & APT75 & 3147.21 & 4.4 & SAAO75 \\
2724.69 & 6.8 & APT75 & 2762.64 & 5.5 & APT75 & 3075.69 & 1.9 & APT75 & 3147.64 & 3.8 & APT75 \\
2725.76 & 5.0 & APT75 & 2762.88 & 5.6 & SSO60 & 3079.71 & 6.8 & APT75 & 3148.64 & 3.7 & APT75 \\
2726.75 & 5.3 & APT75 & 2763.36 & 3.6 & OSN90 & 3080.72 & 6.5 & APT75 & 3149.65 & 3.7 & APT75 \\
2727.66 & 6.7 & APT75 & 2764.65 & 4.8 & APT75 & 3081.68 & 7.6 & APT75 & 3152.20 & 4.1 & SAAO75 \\
2729.64 & 7.6 & APT75 & 2765.65 & 4.0 & APT75 & 3082.67 & 7.7 & APT75 & 3153.65 & 3.3 & APT75 \\
2729.95 & 1.5 & SSO60 & 2766.75 & 2.5 & APT75 & 3086.74 & 4.8 & APT75 & 3155.20 & 3.1 & SAAO75 \\
2730.64 & 7.6 & APT75 & 2768.03 & 0.6 & SSO60 & 3087.66 & 7.6 & APT75 & 3156.19 & 4.0 & SAAO75 \\
2733.64 & 7.4 & APT75 & 2768.65 & 4.9 & APT75 & 3088.79 & 4.6 & APT75 & 3157.30 & 1.5 & SAAO75 \\
2734.63 & 7.6 & APT75 & 2769.36 & 3.6 & OSN90 & 3090.84 & 0.8 & APT75 & 3160.65 & 1.7 & APT75 \\
2735.63 & 7.5 & APT75 & 2769.89 & 1.7 & SSO60 & 3091.65 & 3.9 & APT75 & 3161.65 & 2.7 & APT75 \\
2735.99 & 1.2 & SSO60 & 2770.36 & 2.3 & OSN90 & 3092.66 & 7.3 & APT75 & 3162.65 & 2.7 & APT75 \\
2736.63 & 7.4 & APT75 & 2771.08 & 1.2 & SSO60 & 3093.64 & 7.5 & APT75 & 3163.65 & 2.1 & APT75 \\
2736.94 & 1.0 & SSO60 & 2775.65 & 4.1 & APT75 & 3094.66 & 7.1 & APT75 & 3165.65 & 2.4 & APT75 \\
2737.63 & 7.3 & APT75 & 2776.65 & 4.5 & APT75 & 3095.37 & 2.4 & SAAO50 & 3166.65 & 2.3 & APT75 \\
2738.64 & 4.1 & APT75 & 2777.75 & 1.9 & APT75 & 3101.24 & 6.8 & SAAO50 & 3167.65 & 2.5 & APT75 \\
2740.65 & 6.7 & APT75 & 2778.65 & 4.0 & APT75 & 3102.24 & 6.8 & SAAO50 & 3168.65 & 2.5 & APT75 \\
2743.99 & 4.4 & SSO60 & 2779.66 & 4.0 & APT75 & 3102.73 & 4.9 & APT75 & 3171.65 & 2.2 & APT75 \\
2744.28 & 5.1 & SAAO50 & 2781.65 & 4.1 & APT75 & 3103.63 & 7.3 & APT75 & 3172.65 & 2.2 & APT75 \\
2745.26 & 1.8 & SAAO50 & 2784.67 & 3.3 & APT75 & 3104.62 & 0.7 & APT75 & 3173.65 & 2.1 & APT75 \\
2745.90 & 4.6 & SSO60 & 2785.65 & 3.8 & APT75 & 3106.26 & 6.1 & SAAO50 & 3174.65 & 1.8 & APT75 \\
2747.10 & 1.7 & SSO60 & 2785.90 & 3.2 & SSO60 & 3107.28 & 5.8 & SAAO50 & 3175.65 & 1.9 & APT75 \\
2747.66 & 6.1 & APT75 & 2786.67 & 2.5 & APT75 & 3107.65 & 6.6 & APT75 & 3177.65 & 1.8 & APT75 \\
2747.94 & 5.3 & SSO60 & 2787.66 & 1.9 & APT75 & 3108.24 & 6.7 & SAAO50 & 3187.65 & 1.0 & APT75 \\
\noalign{\smallskip}
\hline
\end{tabular}
\newline
\end{flushleft}
\end{table*}

The resulting light curves of FG Vir are shown in Figs. 1 and 2, where the observations are also
compared with the fit to be derived in the next section.

\begin{figure*}
\centering
\includegraphics*[bb=21 54 568 802,width=170mm,clip]{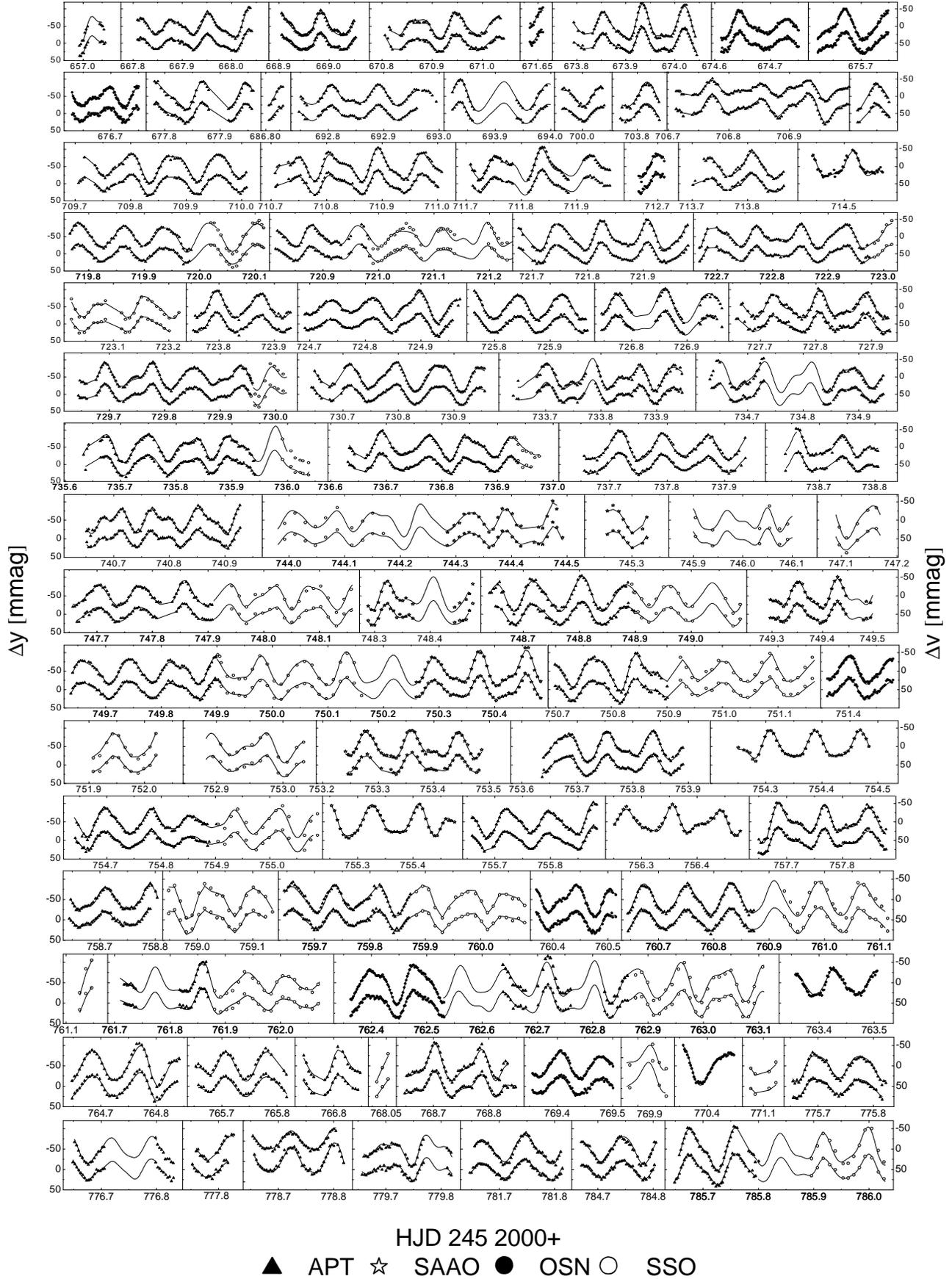}
\caption{Multisite photoelectric three-star-photometry of FG~Vir obtained
during the 2003 and 2004 DSN campaigns. $\Delta y$ and $\Delta v$ are the observed magnitude
differences (variable -- comparison stars) normalized to zero in the narrowband
$uvby$ system. The fit of the 79-frequency solution derived in this paper is
shown as a solid curve. Note the excellent agreement between the measurements
and the fit}
\end{figure*}

\begin{figure*}
\centering
\includegraphics*[bb=21 54 568 800,width=170mm,clip]{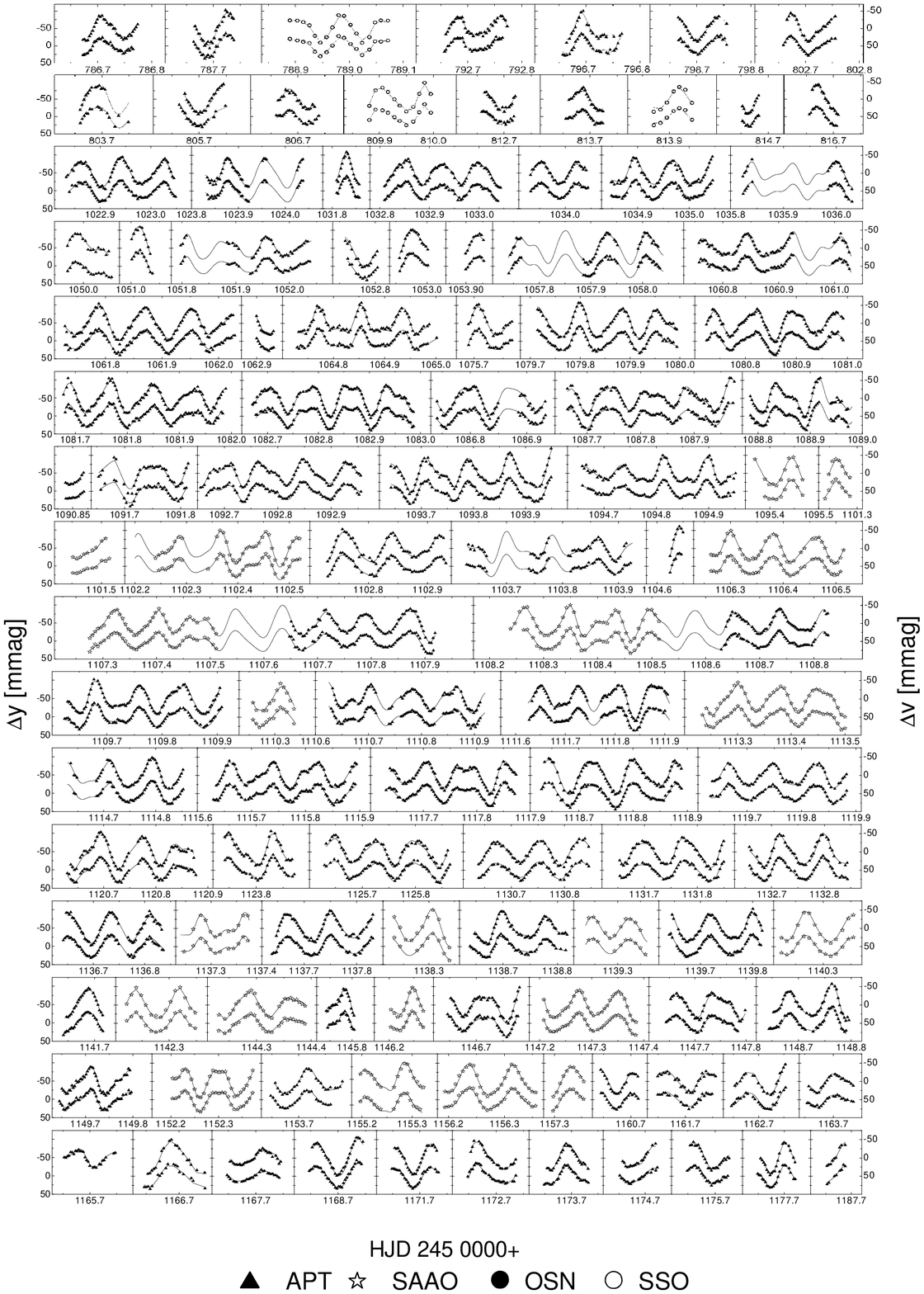}
\caption{Multisite photoelectric three-star-photometry of FG~Vir obtained
during the 2003 and 2004 DSN campaigns, continued}
\end{figure*}

\section{Multiple frequency analysis}

The pulsation frequency analyses were performed with a package of computer
programs with single-frequency and multiple-frequency techniques (PERIOD04,
Lenz \& Breger 2005; http://www.astro.univie.ac.at/$^{\sim}$dsn/dsn/Period04/),
which utilize Fourier as well as multiple-least-squares algorithms. The latter technique fits up to several
hundreds of simultaneous sinusoidal variations in the magnitude domain and does not rely
on sequential prewhitening. The amplitudes and phases of all modes/frequencies are determined by minimizing
the residuals between the measurements and the fit. The frequencies can also be improved at the same time.

Our analysis consists of two parts: We first examine the extensive 2002 - 2004 data and
then add the available 1992 - 1996 data.

\begin{figure*}
\centering
\includegraphics*[width=175mm, clip]{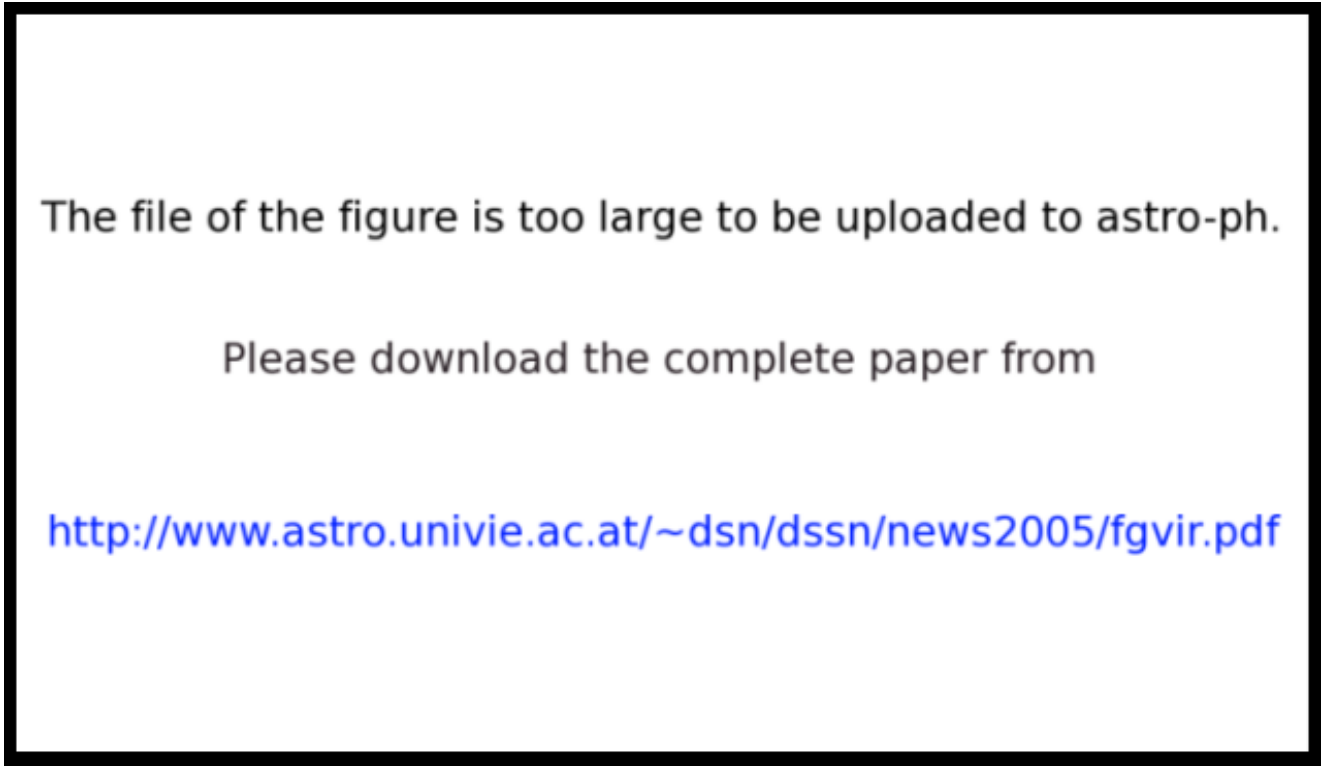}
\caption{Example of power spectra of the 1992--2004 data. Top: Spectral windows showing
effects of the daily and annual aliasing. Bottom: New frequencies detected in the
most difficult frequency region with the lowest amplitudes. The diagram shows that
in the 40--45 c/d region pulsation modes are present and have been detected. The choice of which peaks are
statistically significant depends somewhat on the details of the analysis.}
\end{figure*}

\subsection{Frequencies detected in the 2002 - 2004 data}

The following approach was used in an iterative way:

(i) The data were divided into two data sets to separate the $y$ and $v$ filters, each covering the
total time period from 2002 - 2004. This is necessary because the amplitudes and phasing of the
pulsation are strongly wavelength dependent. In principle, the different amplitudes
could be compensated for by multiplying the $v$ data by an experimentally determined factor
of 0.70 and increasing the weights of the scaled $v$ data accordingly.  (Anticipating the results
presented later in Table 2, we note that this ratio is confirmed by the average amplitude ratio of
the eight modes with highest amplitudes.) However, the small
phase shifts of a few degrees cannot be neglected for the larger-amplitude modes. Consequently,
the data were analyzed together for exploratory analysis, but not for the final analyses.

(ii) We started with the single-frequency solution for the two data sets using the program
PERIOD04. For the Fourier analyses
the two data sets were combined to decrease the noise, while for the actual fits to the data,
separate solutions were made.

(iii) A Fourier analysis was performed to look for additional frequencies/modes from the
combined residuals of the previous solutions. Additional frequencies were then identified and their
signal/noise ratio calculated. Following Breger et al. (1993), a significance criterion of amplitude
signal/noise = 4.0 (which corresponds to power signal/noise of $\sim$12.6) was adopted for
non-combination frequencies. The most clearly detected additional frequencies
were included in a new multifrequency solution. In order for
a new frequency to be accepted as real, the signal/noise criterion also had to be fulfilled
in the multifrequency solution. This avoids false detections due to spill-over effects because
the Fourier technique is a single-frequency technique. Furthermore, since there exist regular
annual gaps, trial annual alias values (separated by 0.0027 c/d) were also examined. We note that
the choice of an incorrect annual alias value usually has little or no effect on the subsequent
search for other frequencies. The choice of an incorrect daily alias (separated by 1 c/d) would
be more serious and we carefully examined different frequency values.

(iv) The previous step was repeated adding further frequencies until no significant
frequencies were found. Note that only the Fourier analyses assume prewhitening; the multiple-frequency solutions
do not.

In this paper we omit the presentation of very lengthy diagrams showing the sequential
detection of new frequencies, except for the example shown in the next subsection. A detailed
presentation of our approach and its results can be found in our analysis of the 2002 data (Breger et al. 2004).

FG~Vir contains one dominant frequency: 12.7162 c/d with a photometric amplitude five times higher
than that of the next strongest mode. To avoid potential problems caused by even small amplitude variability,
for this frequency we calculated amplitudes on an annual basis. The results of the multifrequency analysis
are shown in Table~2. The numbering scheme of the frequencies corresponds to the order of detection,
i.e., the amplitude signal/noise ratio, and therefore differs from that used in
previous papers on FG~Vir.

\subsection{Further frequencies detected in the 1992 - 2004 data}

An extensive photometric data set covering 13 years is essentially unique in
the study of $\delta$~Scuti stars, promising new limits in frequency resolution
and noise reduction in Fourier space. The noise reduction is especially visible
at high frequencies, where the effects of systematic observational errors are small.
The analysis of the combined data had to work around two problems:
the earlier data is not as extensive as the
2002 - 2004 data and there exists a large time gap between 1996 and 2002.

The time gap did lead to occasional uniqueness problems for the frequencies with
amplitudes in the 0.2 mmag range: next to the annual aliasing of 0.0027 c/d we
find peaks spaced 0.00026 c/d, corresponding to a $\sim$ 10 year spacing (see Fig. 3, top right).
Fortunately, the excellent coverage from 2002 - 2004 minimized these ambiguities.

The relatively short coverage of the data from 1992 and 1996 excluded the computation
of 79-frequency solutions for individual years (to avoid overinterpretation).
We have consequently combined
all the $y$ measurements from 1992 to 1996 as well as the available 1995 and 1996 $v$
data. Together with the $y$ and $v$ data sets from 2002 - 2004, we had four data sets.

Fig. 3 illustrates some examples of the resulting power spectra. Due to the large amount of APT data from 2002 - 2004, which
therefore dominates, the 1 c/d aliases are not zero (top left). Nevertheless, due to the excellent
frequency resolution, these aliases are very narrow so that the aliasing problem is not severe.
The figure also shows the power spectrum of the measurements in the 40 - 45 c/d region, which was the most difficult
region for us to analyze due to the small amplitudes of all the detected frequencies.

The new detections are included in Table 2.

\begin{table*}
\caption[]{Frequencies of FG Vir from 2002 to 2004}
\begin{flushleft}
\begin{tabular}{lccccclccccc}
\hline
\noalign{\smallskip}
\multicolumn{3}{c}{Frequency} & Detection$^{(1)}$&  \multicolumn{2}{c}{Amplitude}& \multicolumn{3}{c}{Frequency} & Detection&  \multicolumn{2}{c}{Amplitude}\\
 cd$^{-1}$ &Name& Type & Amplitude & $v$ filter & $y$ filter & \hspace{10mm}cd$^{-1}$ &Name& Type & Amplitude & $v$ filter & $y$ filter \\
&&& S/N ratio & \multicolumn{2}{c}{millimag} &&&& S/N ratio & \multicolumn{2}{c}{millimag} \\
\noalign{\smallskip}
\hline
& & & & $\pm$ 0.04 & $\pm 0.05$ & & & & & $\pm$ 0.04 & $\pm 0.05$\\
5.7491 & f$_ { 47 }$ & & 5.7 & 0.48 & 0.28 & \hspace{10mm}23.3974 & f$_ { 12 }$ & $^{(2)}$ & 22& 1.73 & 1.25 \\
7.9942 & f$_ { 44 }$ & & 5.9 & 0.42 & 0.34 &\hspace{10mm}23.4034 & f$_ { 4 }$ &$^{(2)}$ & 71 & 5.65 & 4.02 \\
8.3353 & f$_ { 78 }$ &f$_6$-f$_1$& $\it{3.8}$ & 0.33 & 0.17 & \hspace{10mm}23.4258 & f$_ { 53 }$ & & 4.9 & 0.37 & 0.28 \\
9.1991 & f$_ { 7 }$ & & 53 & 3.82 & 2.78 & \hspace{10mm}23.4389 & f$_ { 24 }$ & & 9.0 & 0.68 & 0.52 \\
9.6563 & f$_ { 5 }$ & & 71 & 5.20 & 3.61 & \hspace{10mm}23.8074 & f$_ { 48 }$ & & 5.5 & 0.45 & 0.32 \\
10.1687 & f$_ { 25 }$ & & 8.6 & 0.64 & 0.42 & \hspace{10mm}24.0040 & f$_ { 58 }$ & & 4.6 & 0.35 & 0.31 \\
10.6872 & f$_ { 79 }$ &f$_4$-f$_1$& 3.5 & 0.26 & 0.15 & \hspace{10mm}24.1940 & f$_ { 10 }$ & & 29 & 2.26 & 1.58 \\
11.1034 & f$_ { 20 }$ & & 11 & 0.67 & 0.65 & \hspace{10mm}24.2280 & f$_ { 3 }$ & & 74 & 5.79 & 4.20\\
11.2098 & f$_ { 38 }$ & & 6.4 & 0.35 & 0.43 & \hspace{10mm}24.3485 & f$_ { 18 }$ & & 12 & 0.99 & 0.63\\
11.5117 & f$_ { 70 }$ &f$_3$-f$_1$& 4.0 & 0.30 & 0.18 & \hspace{10mm}24.8703 & f$_ { 36 }$ &f$_1$+f$_2$ & 6.4 & 0.48 & 0.38 \\
11.6114 & f$_ { 50 }$ & & 5.3 & 0.35 & 0.27 & \hspace{10mm}25.1788 & f$_ { 30 }$ & & 7.3 & 0.59 & 0.39 \\
11.7016 & f$_ { 33 }$ & & 6.9 & 0.42 & 0.44 & \hspace{10mm}25.3793 & f$_ { 45 }$ & & 5.7 & 0.37 & 0.31 \\
11.8755 & f$_ { 63 }$ & & 4.4 & 0.27 & 0.25 & \hspace{10mm}25.4324 & f$_ { 15 }$ &2f$_1$ & 16 & 1.23 & 0.89 \\
11.9421 & f$_ { 29 }$ & & 7.6 & 0.55 & 0.38 & \hspace{10mm}25.6387 & f$_ { 68 }$ & & 4.0 & 0.30 & 0.26 \\
12.1541 & f$_ { 2 }$ &$^{(2)}$ & 85 & 6.09 & 4.21 & \hspace{10mm}26.5266 & f$_ { 71 }$ & & $\it{4.7}$& 0.30 & 0.20 \\
12.1619 & f$_ { 14 }$ &$^{(2)}$ & 16 & 1.13 & 0.83 & \hspace{10mm}26.8929 & f$_ { 61 }$ & & 4.5 & 0.34 & 0.27 \\
12.2158 & f$_ { 52 }$ & & 5.1 & 0.38 & 0.24 & \hspace{10mm}26.9094 & f$_ { 75 }$ & & $\it{4.2}$& 0.34 & 0.19 \\
12.7162 & f$_ { 1 }$ & & 442 & 31.74 & 21.92 & \hspace{10mm}28.1359 & f$_ { 19 }$ & & 11 & 0.87 & 0.55\\
12.7944 & f$_ { 17 }$ & & 13 & 0.88 & 0.66 & \hspace{10mm}29.4869 & f$_ { 51 }$ &f$_7$+f$_{11}$ & 5.2 & 0.39 & 0.28\\
13.2365 & f$_ { 27 }$ & & 8.3 & 0.45 & 0.56 & \hspace{10mm}30.9146 & f$_ { 60 }$ & & 4.5 & 0.31 & 0.27 \\
14.7354 & f$_ { 49 }$ & & 5.3 & 0.29 & 0.38 & \hspace{10mm}31.1955 & f$_ { 57 }$ & & 4.6 & 0.35 & 0.27 \\
16.0711 & f$_ { 13 }$ & & 20 & 1.56 & 1.07 & \hspace{10mm}31.9307 & f$_ { 34 }$ & & 6.6 & 0.49 & 0.40 \\
16.0909 & f$_ { 31 }$ & & 7.3 & 0.55 & 0.39 & \hspace{10mm}32.1895 & f$_ { 32 }$ & & 7.0 & 0.56 & 0.38 \\
19.1642 & f$_ { 26 }$ & & 8.6 & 0.55 & 0.59 & \hspace{10mm}33.0437 & f$_ { 74 }$ &$^{(4)}$& $\it{4.3}$& 0.29 & 0.19 \\
19.2278 & f$_ { 9 }$ & & 30 & 2.51 & 1.69 & \hspace{10mm}33.7677 & f$_ { 43 }$ &f$_1$+f$_6$ & 5.9 & 0.44 & 0.31 \\
19.3259 & f$_ { 41 }$ & & 6.3 & 0.53 & 0.34 & \hspace{10mm}34.1151 & f$_ { 22 }$ & & 9.8 & 0.75 & 0.49 \\
19.6439 & f$_ { 65 }$ & & 4.3 & 0.40 & 0.22 & \hspace{10mm}34.1192 & f$_ { 72 }$ & & $\it{4.7}$ & 0.21 & 0.25 \\
19.8679 & f$_ { 8 }$ &$^{(3)}$& 55 & 4.44 & 3.19 & \hspace{10mm}34.1864 & f$_ { 54 }$ & & 4.9 & 0.37 & 0.25 \\
19.8680 &   &$^{(3)}$& 30 & 2.40 & 1.78 & \hspace{10mm}34.3946 & f$_ { 55 }$ & & 4.6 & 0.31 & 0.28 \\
20.2878 & f$_ { 11 }$ & & 26 & 2.13 & 1.45 & \hspace{10mm}34.5737 & f$_ { 23 }$ & & 9.3 & 0.69 & 0.45 \\
20.2925 & f$_ { 56 }$ & & 4.6 & 0.31 & 0.39 & \hspace{10mm}35.8858 & f$_ { 76 }$ & & $\it{4.1}$& 0.22 & 0.21 \\
20.5112 & f$_ { 35 }$ & & 6.6 & 0.41 & 0.52 & \hspace{10mm}36.1196 & f$_ { 40 }$ &f$_1$+f$_4$& 6.3 & 0.40 & 0.31 \\
20.8348 & f$_ { 39 }$ & & 6.3 & 0.53 & 0.38 & \hspace{10mm}36.9442 & f$_ { 37 }$ &f$_1$+f$_3$& 6.4 & 0.43 & 0.27 \\
21.0515 & f$_ { 6 }$ & & 55 & 4.43 & 3.08 & \hspace{10mm}39.2165 & f$_ { 69 }$ & & 4.0 & 0.27 & 0.16 \\
21.2323 & f$_ { 16 }$ & & 14 & 1.06 & 0.80 & \hspace{10mm}39.5156 & f$_ { 59 }$ &f$_{9}$+f$_{11}$& 4.5 & 0.24 & 0.25 \\
21.4004 & f$_ { 46 }$ & & 5.7 & 0.52 & 0.33 & \hspace{10mm}42.1030 & f$_ { 64 }$ &2f$_6$ & 4.3 & 0.22 & 0.28 \\
21.5507 & f$_ { 28 }$ & & 7.9 & 0.61 & 0.42 &\hspace{10mm}42.1094 & f$_ { 62 }$ & & 4.5 & 0.24 & 0.25 \\
22.3725 & f$_ { 42 }$ &f$_1$+f$_5$ & 6.2 & 0.52 & 0.35 & \hspace{10mm}43.0134 & f$_ { 73 }$ & & $\it{4.4}$& 0.21 & 0.15 \\
23.0253 & f$_ { 66 }$ & & 4.1 & 0.30 & 0.23 & \hspace{10mm}43.9651 & f$_ { 67 }$ & & 4.0 & 0.26 & 0.17 \\
23.3943 & f$_ { 21 }$ & & 11 & 0.85 & 0.60 & \hspace{10mm}44.2591 & f$_ { 77 }$ & & $\it{4.0}$ & 0.20 & 0.18 \\
\noalign{\smallskip}
\hline
\end{tabular}
\newline
$^{(1)}$ The noise for the amplitude signal/noise ratios were calculated over a 4 cd$^{-1}$ range. Limits for a significant detection
are 4.0 for independent frequencies and 3.5 for combination modes with known values. Numbers in italics
indicate 1992 - 2004 data (see text).\\
$^{(2)}$ The close frequencies, 12.1541 and 12.1619 as well as 23.3974 and 23.4034 c/d, are all separate modes.
In short data sets this could lead to an erroneous
identification as single modes with variable amplitude.\\
$^{(3)}$ For the possible frequency pair near 19.868 c/d the existence of two separate modes cannot be proved at this point.
A single frequency with a slowly variable amplitude (beat period $\sim$21.5 years) is also possible.\\
$^{(4)}$ The 2002--2004 data clearly show a mode at 33.044 c/d, though with
considerably reduced amplitudes from 1992-1995 data. Breger et al. (1998) listed the value of the frequency
as 33.056 c/d, which was the highest peak from a broad selection of peaks separated by annual aliases 0.0027 c/d apart.
We note that in the new data, a value separated by 1 annual alias, viz., 33.0461 c/d, is also possible.\\

\end{flushleft}
\end{table*}

A comparison with the frequencies published in earlier papers shows that all the previously
detected frequencies were confirmed. This also includes those previously detected modes not found to be
statistically significant in the 2002 data alone. In a few cases, different $annual$ aliases were selected.
However, the main result is the increase in the number of detected frequencies to 79, which more than doubles
the previous results.

Fig.~4 shows the distribution of frequencies in frequency space. We note the wide range of excited frequencies,
which is unusual for $\delta$~Scuti stars, as well as the clustering of the excited frequencies. This clustering
persists even after the suspected combination frequencies and 2f harmonics are removed.

A new feature is the detection of frequencies with values between 40 and 45 c/d. They all have small amplitudes
of $\Delta y\sim$0.2 mmag. The lower noise of the new data now made their detection possible.

\begin{figure*}
\centering
\includegraphics*[bb=44 329 790 768, width=175mm, clip]{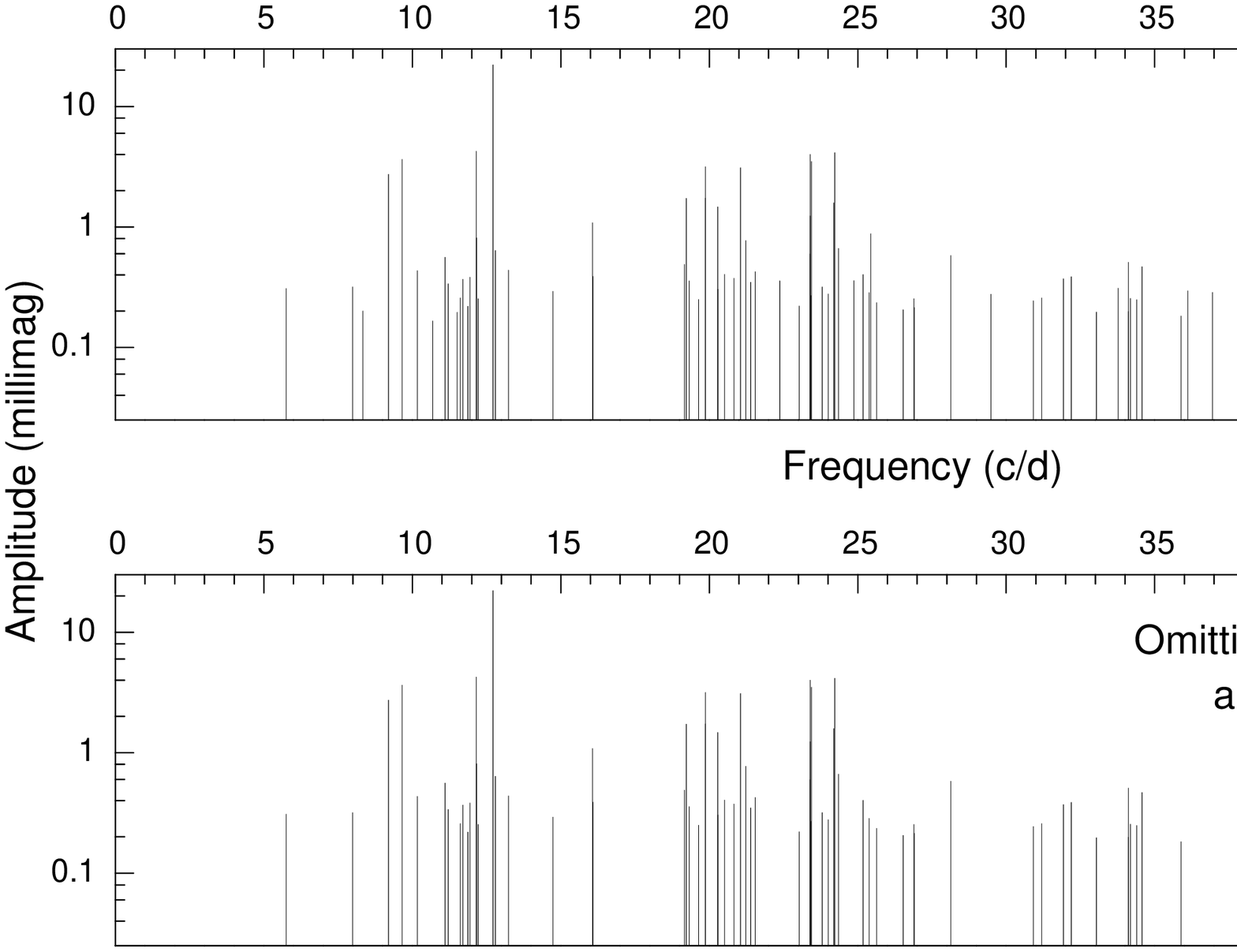}
\caption{Distribution of the frequencies of the detected modes. The diagram suggests that
the excited pulsation modes are not equally distributed in frequency.}
\end{figure*}

\subsection{Color effects}
\begin{table*}
\caption{Phase differences and amplitude ratios}
\begin{flushleft}
\begin{tabular}{lccccc}
\hline
\noalign{\smallskip}
\multicolumn{2}{c}{Frequency}& \multicolumn{2}{c}{Phase differences
in degrees} & \multicolumn{2}{c}{ Amplitude ratios}\\
& c/d &  \multicolumn{2}{c}{$\phi_v - \phi_y$} & \multicolumn{2}{c}{$v/y$}\\
& & 2002--2004 & 1995--2004$^{1)}$ & 2002--2004 & 1995--2004\\
\noalign{\smallskip}
\hline
\noalign{\smallskip}
$f_1$ & 12.716 &  --1.7 $\pm$ 0.1 & --1.5  & 1.45 $\pm$ 0.00 &  1.45\\
$f_2$ & 12.154 &  +3.1 $\pm$ 0.7 & +2.8 & 1.44 $\pm$ 0.02 & 1.44\\
$f_3$ & 24.227 & --3.0 $\pm$ 0.7 & --3.0 & 1.38 $\pm$ 0.02 & 1.40  \\
$f_4$ & 23.403 & --3.5 $\pm$ 0.7 & --3.3 & 1.40 $\pm$ 0.02 & 1.42\\
$f_5$ & 9.656 &  --5.1 $\pm$ 0.8 & --4.5  & 1.44 $\pm$ 0.02 & 1.43 \\
$f_6$ & 21.051 &  --3.7 $\pm$ 1.0  & -4.2  & 1.44 $\pm$ 0.02 & 1.45\\
$f_7$ & 9.199 & --6.7 $\pm$ 1.1 & --6.6  & 1.38 $\pm$ 0.03 & 1.40\\
$f_8$ & 19.868$^{2)}$ &  --4.3 $\pm$ 1.9 & --2.9 & 1.45 $\pm$ 0.05 & 1.44 \\
$f_9$ & 19.228 & --4.6 $\pm$ 1.7 & --3.9  & 1.48 $\pm$ 0.05 & 1.46  \\
$f_{10}$ & 24.194 & --1.2 $\pm$ 1.9 & -0.6 & 1.43 $\pm$ 0.07 & 1.38 \\
\noalign{\smallskip}
%\multicolumn{3}{l}{Number of hours $y/v$}& 412/292 & 494/374\\
%\noalign{\smallskip}
\hline
\end{tabular}\newline
$^{1)}$ Error estimates omitted: usually lower than for 2002--2004, but some instability
is possible due to large time gap\\
$^{2)}$ Using single frequency with annual amplitude variations
\end{flushleft}
\end{table*}

The light curves of pulsating stars are not identical at different wavelengths.
In fact, amplitude ratios and phase shifts provide a tool for the identification
of nonradial modes (e.g., see Garrido et al. 1990, Moya et al. 2004).
For $\delta$ Scuti stars, the amplitude ratios between different colors are primarily
dependent on the surface temperature. For the individual pulsation modes, the phase differences
and deviations from the $average$ amplitude ratio are small. This means that observational
errors need to be small and any systematic errors between the different colors should be avoided.

For most nights there exist both $v$ and $y$ passband data, so that amplitude ratios
as well as phase differences can be derived. However, our 79-frequency solution is not
perfect. In order not to introduce systematic errors in the phase differences and amplitude
ratios, for the calculation of amplitude ratios and phase differences, 
we have omitted those nights for which two-color data are not available. Consequently,
no data from 1992 and 1993 were used and all 1995 (single-color) CCD measurements were omitted.

Table 3 lists the derived phase differences and amplitude ratios for the modes with
relatively high amplitudes. The uncertainties listed were derived from error-propagation
calculations based on the standard formulae given by Breger et al. (1999). The results
can now be used together with spectroscopic line-profile analyses to identify the pulsation modes.

\begin{figure*}
\centering
\includegraphics*[bb=80 439 790 775, width=175mm, clip]{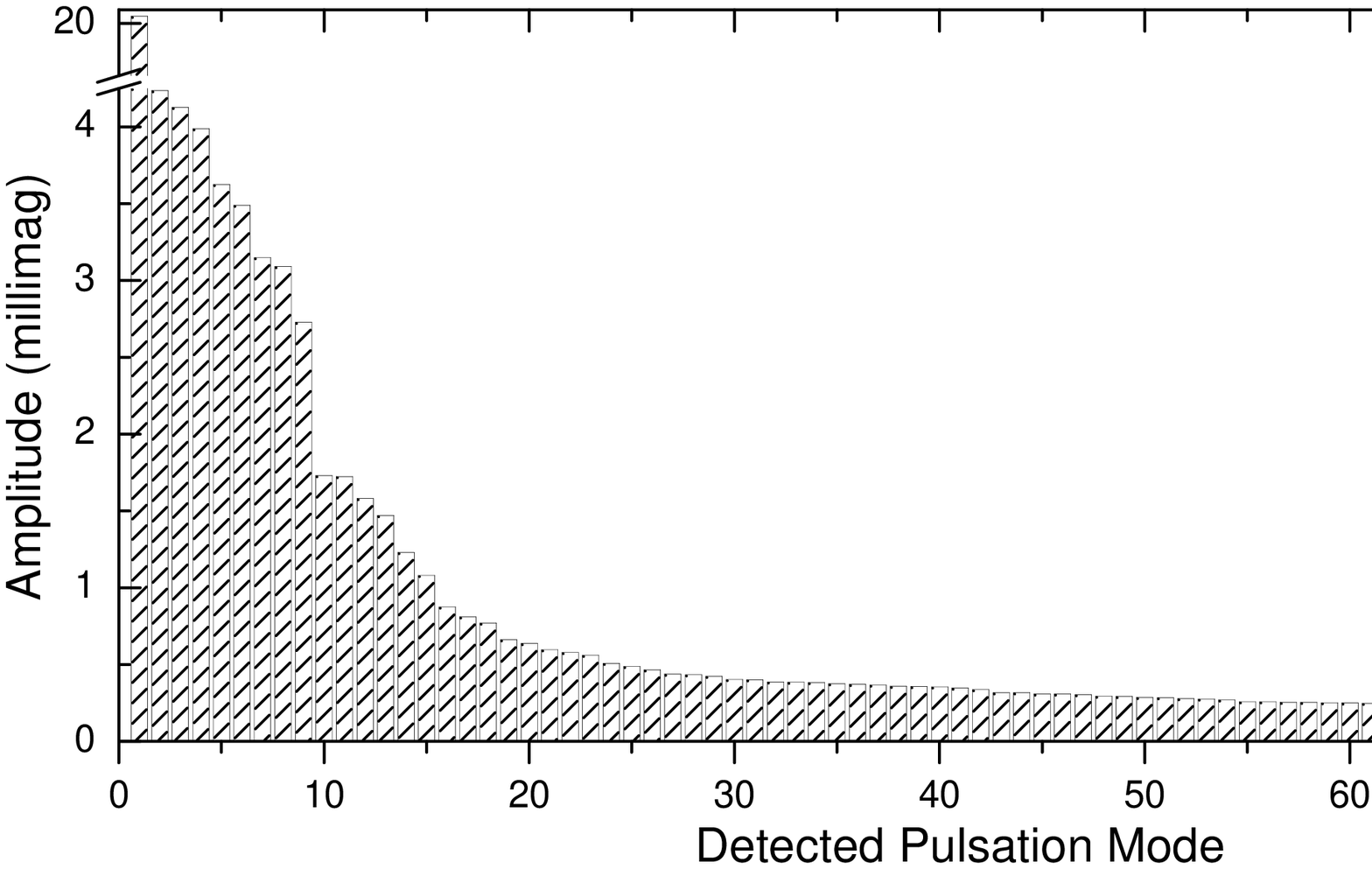}
\caption{Distribution of the amplitudes of the frequencies
with significant detections. To increase the accuracy, we have computed
amplitudes from 0.5*($y$ amplitude + 0.70 $v$ amplitude) to simulate the
amplitude in the $y$ passband. Note the large number of detected modes with amplitudes
near the detection limit of 0.2 mmag. This suggests that even moderate increases in the
amount of data lead to considerably higher number of detections.}
\end{figure*}

\begin{figure*}
\centering
\includegraphics*[bb=46 401 790 752, width=175mm, clip]{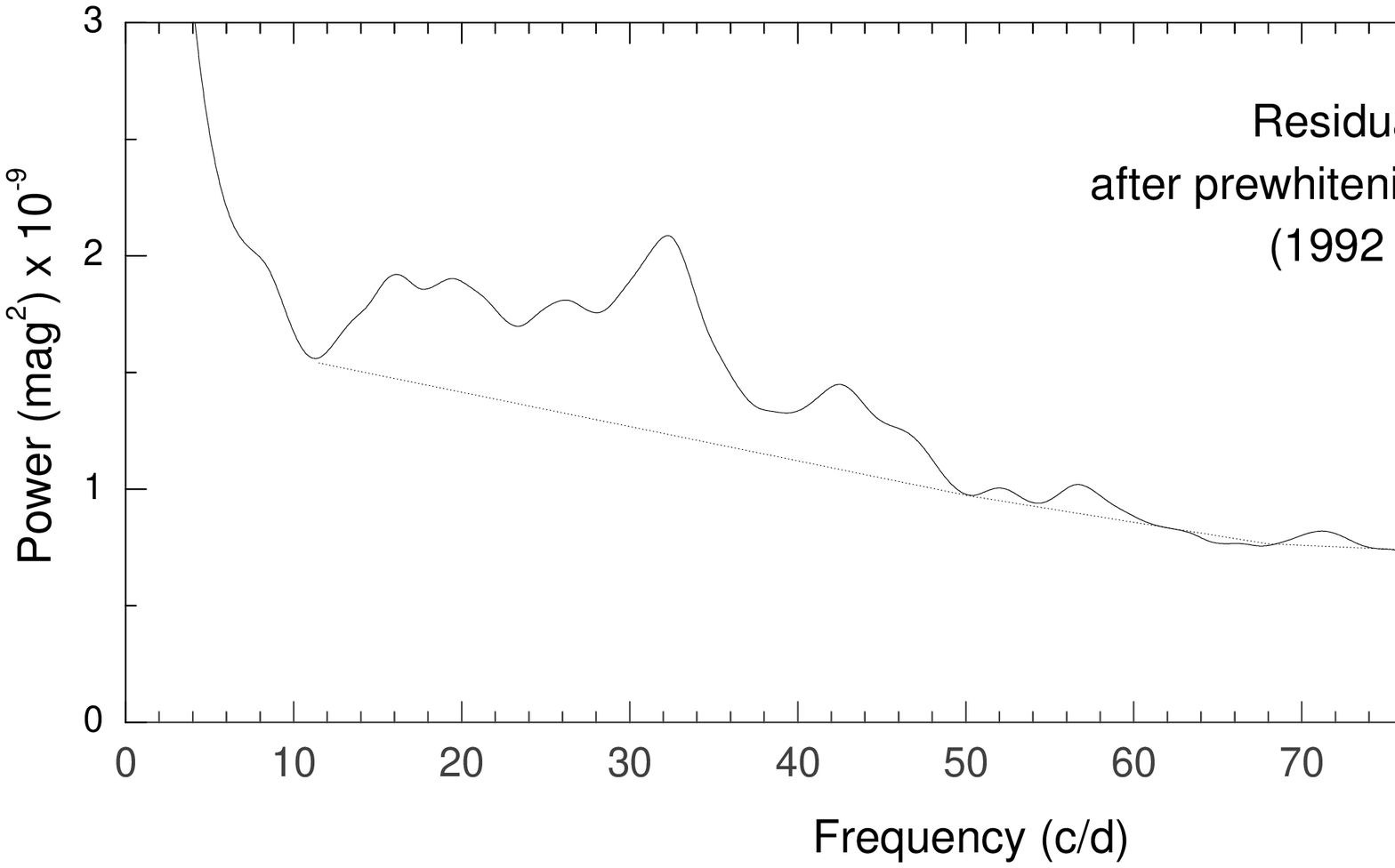}
\caption{Power of the residual noise of the 1992 -- 2004 data after subtracting 80 (79) frequencies.
(The additional frequency is the close doublet at 19.868 c/d adopted for the long time span.)
The average amplitude in the amplitude spectrum was calculated for 2 c/d regions and squared to
give power. The strong increase towards lower frequencies is caused by observational noise and
standard in terrestrial photometry. Note the excess power in 10 - 50 c/d region, shown
above the dotted curve. This shows that many additional pulsation modes exist in
the same frequency region in which the already detected frequencies occur.}
\end{figure*}

\section{Combination frequencies}

We have written a simple program to test which of the 79 frequencies found can be
expressed as the sum or differences of other frequencies. Due to the excellent frequency
resolution of the 2002 - 2004 data, we could be very restrictive in the identification
of these combinations. A generous limit of $\pm$ 0.001 c/d was adopted. The probability of incorrect identifications
is correspondingly small. A number of combinations was found and these
are marked in Table~2. They generally agreed to $\pm$ 0.0002 c/d.

How many accidental agreements do we expect? We have calculated this number through a large number of
numerical simulations, assuming a reasonable agreement of the observed frequency to within 0.0002 c/d
of the predicted frequency. We obtain an average of 0.93 accidental matchings of peaks with combination frequencies.
We conclude that most or all detected combination frequencies are not accidental. The argument is strengthened by the fact that the
combinations detected by us all contain one of two specific modes, which reduces the chance of accidental
agreements to essentially zero.

We also note that the lowest frequency detected, f$_{47}$ at 5.749 c/d, can be expressed as
a $triple$ combination of $f_1$, f$_3$ and f$_{57}$. This may be accidental.

\subsection{Combinations of the dominant mode at 12.7163 c/d}

Due to the presence of a dominant mode at 12.7163 c/d (f$_1$), it is not surprising that some
combination frequencies, f$_1$$\pm$f$_i$, exist and are detected (see Table 2). In order to
examine this further, we have performed additional calculations with the 2002 - 2004 data. We have repeated
our multifrequency solutions described earlier while omitting all frequency combinations of f$_1$. The
residuals of the $y$ and $v$ data were combined to form the sum ($y$ + 0.70$v$) to account for the different
amplitudes at the two passbands. A new multifrequency solution containing the possible combinations of
the dominant mode with f$_2$ through f$_8$ was made. The results are shown in Table 4, in which small differences
compared to Table 2 are caused by the different procedures of our analyses. 
The amplitudes of the sums (f$_1$+f$_i$) are higher than those of the differences (f$_1$-f$_i$). We
believe the result to be real and intrinsic to the star. The differences are generally found at low frequencies,
where the observational noise is higher. Increased noise should lead to higher amplitudes, which are not found.

\begin{table}
\caption[]{Combination frequencies involving the dominant mode}
\begin{flushleft}
\begin{tabular}{lcc}
\hline
\noalign{\smallskip}
f$_1 \pm$ & \multicolumn{2}{c}{Amplitude in $y$ (2002--2004)}\\
& Sum of frequencies & Difference of frequencies\\
& mmag & mmag\\
\noalign{\smallskip}
\hline
\noalign{\smallskip}
f$_2$ & 0.35 & (0.06) \\
f$_3$ & 0.28 & 0.19\\
f$_4$ & 0.31 & 0.17 \\
f$_5$ & 0.33 & (0.16)\\
f$_6$ & 0.39 & 0.21 \\
f$_7$ & (0.10) & (0.07)\\
f$_8$ & (0.07) & (0.13)\\
\noalign{\smallskip}
\hline
\noalign{\smallskip}
\end{tabular}
%\noalign{\smallskip}
\newline
The amplitudes in brackets are too low for fulfilling the adopted criterion
of statistical significance.\\
\end{flushleft}
\end{table}

\subsection{Combinations of the 20.2878 c/d mode}

Apart from the mode combinations involving the dominant mode, f$_1$, two other two-mode combinations are found,
both involving f$_{11}$ at 20.2878 c/d. While the mode identifications are still in progress,
this mode can to a high probability be identified as a $\ell$=1, $m$=-1 mode (Zima et al. 2003).
At first sight, this appears surprising, since the photometric amplitude is only 1.5 mmag. However, for the
known inclination and mode identification, we calculate a geometric cancellation factor of $\sim$3.6, so that
the real amplitude of this mode is 5 mmag or larger.

The fact that the $m$ value is not equal to zero has some interesting consequences for the observations of
combination frequencies. The reason is that there are two frames of reference: the stellar frame corotating with the star, and
that of the observer. For nonradial modes of $m$ values $\neq$ 0 (i.e., waves traveling around the star), the
frequencies between the two frames of reference differ by $m\Omega$, where $\Omega$ is the rotation frequency of the
star (see Cox 1984 for an excellent discussion). The frequency combinations occur in the corotating (stellar)
frame, and not the observer's frame of reference.
Consequently, many possible combinations involving non-axisymmetric modes should not be observed as simple sums or differences
of observed frequency values.

It follows that for the non-axisymmetric ($m$=-1) mode at 20.2878 c/d, our simple method to search for combination frequencies
from the observed frequency values may only detect combinations, f$_i$ + f$_j$, with $m$ = +1 modes.

This strict requirement is met by the two identified combinations of 20.2878 c/d! Both coupled modes, f$_9$ = 19.2278 c/d
as well as f$_{11}$ = 20.2878 c/d have been identified as $m$ = +1 modes. Such a combination of $m$ values of opposite sign
can be detected because the combination is invariant to the transformation between the two frames of reference:
$ -m\Omega + m\Omega$ = 0.

\section{The problem of missing frequencies solved?}

$\delta$ Scuti star models predict pulsational instability in many radial and
nonradial modes.  The observed number of low-degree modes is much lower than
the predicted number.  The problem of mode selection is most severe for
post-main-sequence $\delta$~Scuti stars, which comprise about 40 percent of the
observed $\delta$~Scuti stars. The theoretical frequency spectrum of unstable
modes is very dense. Most modes are of mixed character: they behave like
p-modes in the envelope and like g-modes in the interior.  For example, for a
model of the relatively evolved star 4 CVn, the models predict 554 unstable
modes of $\ell$ = 0 to 2, i.e., 6 for $\ell$ =
0, 168 for $\ell$ = 1, and 380 for $\ell$ = 2 (see Breger \& Pamyatnykh 2002).
However, only 18 (and additional 16 combination frequencies) were observed (Breger et al. 1999).
The problem also exists for other $\delta$~Scuti stars. A complication occurs
since the models so far cannot predict the amplitudes of the excited modes.

Two explanations offer themselves: the missing modes exist, but have amplitudes too small to
have been observed, or there exists a mode selection mechanism, which needs to be examined in
more detail. Promising scenarios involve the selective excitation of modes trapped in the envelope
(Dziembowski \& Kr\'olikowska 1990) or random mode selection.

Let us turn to the star FG Vir. Unpublished models computed by A. A. Pamyatnykh predict 80 unstable modes with
$\ell$ = 0, 1 and 2 in the 8 - 40 c/d range. This number is smaller than that mentioned previously
for the more evolved star 4 CVn, but until now this large number was not observed either.
The present study addresses the question by lowering the observational amplitude threshold to 0.2 mmag. We have
detected 79 frequencies, of which 12 could be identified as harmonics or combination frequencies. This leaves
67 independent frequencies. There also exists considerable evidence that many more modes are excited:

(i) Consider the amplitude distribution of the detected modes shown in Fig. 5. There is a rapid increase in the
number of modes as one goes towards low amplitudes. The present limit near 0.2 mmag is purely observational. Consequently,
the number of excited modes must be much larger.

(ii) Consider the power spectrum of the residuals after subtraction of the multifrequency solution (Fig. 6).
We see excess power in the 10 - 50 c/d range. This is exactly the region in which
the previously detected modes were found. This indicates that many additional modes similar to the ones detected previously
are excited at small amplitude.

We can exclude the possibility that the large number of observed frequencies is erroneous because of
imperfect prewhitening due to amplitude variability. We have examined this possibility in great detail,
with literally thousands of different multifrequency solutions allowing for amplitude variability. In no case was it
possible to significantly reduce the structure in the power spectrum of the residuals. In fact, the 'best' multifrequency
solution adopted treated the two colors as well as the 1992 - 1996 and 2002 - 2004 data separately and
allowed annual amplitude variability of the dominant mode at 12.7162 c/d. Amplitude variability, therefore, cannot
explain the excess power.

{\it{Consequently, the problem that the number of detected modes is much smaller than the
number of predicted low-$\ell$ modes no longer exists, at least for FG Vir.}} Of course, we cannot conclude that each
theoretically predicted mode is really excited and has been detected. This would require much more extensive mode identifications than
are available at this stage.

In the previous discussion, we have concentrated on the low-$\ell$ modes which are easily observed photometrically.
One also has to consider that at low amplitudes, variability from modes of higher $\ell$ values might also be
seen photometrically. The geometric cancellation effects caused by the integration over the whole surface only
become important for $\ell \geq$ 3. This is shown by Daszy\'nska-Daszkiewicz et al. (2002), who calculated the amplitude reduction
factors caused by temperature variations across the disk. From this paper we can roughly estimate a cancellation factor of $\sim$50 in the
$y$ passband for $\ell$ = 3 modes, implying that only the largest-amplitude modes could be photometrically detected. Such modes would
have amplitudes similar to, or larger, than that of the observed dominant ($\ell$ = 1) mode at 12.7162 c/d and might therefore be expected to be few.
Regrettably, the situation is somewhat more complicated. The results presented in Fig. 2 of the  Daszy\'nska-Daszkiewicz et al. paper
are actually based on models with higher surface temperatures and could fit the $\beta$ Cep variables. A.  Pamyatnykh has kindly calculated
specific models fitting the star FG Vir. Here the geometric cancellation factor becomes smaller by a factor of two or three.
Consequently, some of the low-amplitude modes observed by us could also be $\ell$ = 3 modes. 

We conclude that the large number of detected frequencies as well as the large number of additional frequencies suggested by the power spectrum
of the residuals confirms the theoretical prediction of a large number of excited modes. A mode-by-mode check for each predicted
mode is not possible at this stage.

\section*{Acknowledgements}
This investigation has been supported by the
Austrian Fonds zur F\"{o}rderung der wissenschaftlichen Forschung. The Spanish observations
were supported by the Junta de Andalucía and the DGI under project AYA2000-1580. We are grateful
to P. Reegen for help with the APT measurements, M. Viskum for providing a table of the individual
measurements of the 1996 photometry not listed in his paper, A. A. Pamyatnykh and J. Daszy\'nska-Daszkiewicz for preliminary
pulsation models, as well as W. Dziembowski and W. Zima for important discussions.

\end{document}